\documentclass{bvm2022} % Ändern Sie diese Zeile NICHT

\makeatletter
\let\blx@rerun@biber\relax
\makeatother

\begin{document}
% Ändern Sie oberhalb von \begin{document} nichts außer der Angabe der Bibliografie-Datei, jegliche sonstigen Änderungen und Einträge gehen verloren

% Wählen Sie mit dem Befehl \selectlanguage die Sprache aus, in der Ihr Proceeding verfasst ist:
%\selectlanguage{ngerman} % Deutsch 
\selectlanguage{english} % Englisch

% Angabe des Titels Ihres Beitrages
\title{Automatic Classification of Neuromuscular Diseases in Children Using Photoacoustic Imaging}
% Falls Sie einen Kurbeitrag schreiben, muss der Titel mit "Abstract:" starten
% \title{Abstract: Bildverarbeitung für die Medizin 2022}

% Optionale Angabe des Subtitels
%\subtitle{Richtlinien zur Erstellung der druckfertigen Beiträge}

% titlerunning erscheint in der Kopfzeile jeder zweiten Seite
% LaTeX erzeugt diesen automatisch aus Ihrem Beitragstitel
% Ist dieser jedoch zu lang erscheint stattdessen die Nachricht "Title Suppressed Due to Excessive Length"
% Geben Sie in diesem Fall eine Kurzform des Titel hier an
\titlerunning{Photoacoustic Imaging for Neuromuscular Diseases}

% Bitte geben Sie alle beteiligten Autoren an
% Um bei jedem Autor den Nachnamen richtig identifizieren zu können, schreiben Sie diesen jeweils in den Befehl \lname{}
% Bei Beteiligung mehrerer Institute, nennen Sie die Nummer des Instituts bzw. der Institute (siehe weiter unten) mit \inst{} nach dem jeweiligen Autor. Ist nur ein Institut beteiligt, lassen Sie dies weg.
% Trennen Sie alle Autoren durch ein Komma
\author{
	Maja \lname{Schlereth} \inst{1,2}, 
	Daniel \lname{Stromer} \inst{2}, 
	Katharina \lname{Breininger} \inst{1}, 
	Alexandra \lname{Wagner} \inst{3}, 
	Lina \lname{Tan} \inst{3}, 
	Andreas \lname{Maier} \inst{2},
	Ferdinand \lname{Knieling} \inst{3}
}

% Geben Sie hier die Autoren so an, wie Sie in der Kopfzeile erscheinen sollen
% Nennen Sie nur die Nachnamen 
% Folgen Sie abhängig von der Anzhal der beteiligten Autoren den folgenden Beispielen
% \authorrunning{Meier} - ein Autor
% \authorrunning{Meier \& Müller} - zwei Autoren
% \authorrunning{Meier, Müller \& Schulze} - drei Autoren
% \authorrunning{Meier et al.} - mehr als drei Autoren
\authorrunning{Schlereth et al.}

% Geben Sie die beteiligten Institute an
% Bei Beteiligung mehrerer Institute ist jedem Institut eine aufsteigende Nummer mit \inst{} voranzustellen.
% Ist nur ein Institut beteiligt, lassen Sie die Nummer entsprechend weg.
% Trennen Sie einzelne Institute mit \\
\institute{
\inst{1} Department of Artificial Intelligence in Biomedical Engineering, \\FAU Erlangen-N\"urnberg, Erlangen\\
\inst{2} Pattern Recognition Lab, FAU Erlangen-N\"urnberg, Erlangen\\
\inst{3} Department of Pediatrics and Adolescent Medicine, Universit\"atsklinik Erlangen, FAU Erlangen-N\"urnberg, Erlangen
}

% Geben Sie die E-Mail-Adresse des korrespondierenden Autors an
\email{maja.schlereth@fau.de}

\maketitle

% Kurzfassung Ihres Beitrags, nur für Langbeitrage
% Nutzen Sie \begin{abstract} ... \end{abstract} NICHT für Kurbeiträge
\begin{abstract}
Neuromuscular diseases (NMDs) cause a significant burden for both healthcare systems and society. They can lead to severe progressive muscle weakness, muscle degeneration, contracture, deformity and progressive disability. The NMDs evaluated in this study often manifest in early childhood. As subtypes of disease, e.g. Duchenne Muscular Dystropy (DMD) and Spinal Muscular Atrophy (SMA), are difficult to differentiate at the beginning and worsen quickly, fast and reliable differential diagnosis is crucial. Photoacoustic and ultrasound imaging has shown great potential to visualize and quantify the extent of different diseases. The addition of automatic classification of such image data could further improve standard diagnostic procedures. We compare deep learning-based 2-class and 3-class classifiers based on VGG16 for differentiating healthy from diseased muscular tissue. This work shows promising results with high accuracies above $0.86$ for the 3-class problem and can be used as a proof of concept for future approaches for earlier diagnosis and therapeutic monitoring of NMDs.
\end{abstract}

\section{Introduction}
Duchenne Muscular Dystropy (DMD) and Spinal Muscular Atrophy (SMA) often manifest in early childhood and are hard to differentiate in early stages~\cite{ref-1}. In severe cases, the muscles degenerate quickly, and reliable and fast diagnosis is necessary to start with the correct treatment as early as possible. Currently, confirmation of the diagnosis is typically done by genetic testing which takes up to several months~\cite{ref-2}.
To support an early assessment, imaging-based diagnosis has been proposed. Photoacoustic (PA) imaging has been investigated as an imaging modality for assessing neuromuscular diseases~\cite{ref-3}. It comes with many advantages, as it is non-invasive, has a low acquisition time and is able to differentiate tissue composition based on their PA properties. While ultrasound (US) images depict the morphology of different tissue types, PA images contain functional information. For instance, the wavelength 800$\ts$nm helps to observe highly perfused and hemoglobin or myoglobin rich tissue, such as muscular or brain tissue~\cite{ref-4,ref-5}. Regensburger et\,al.\ show that PA imaging can be successfully used to detect collagen tissue as a biomarker for DMD as muscular tissue degenerates in DMD and is replaced by collagen or fatty tissue~\cite{ref-6}. The authors found significant differences between healthy and diseased tissue regarding signal intensity for collagen. To analyse the data, regions of interest were manually drawn and individually evaluated for each patient. To overcome these limitations, automatic image processing and classification can be used~\cite{ref-7}. Zhang et\,al.\ propose a traditional machine learning (ML) and a deep learning (DL) method using AlexNet and GoogLeNet for classification of breast cancer for PA images with high accuracies~\cite{ref-8}, which motivates the use of DL for diagnosis of SMA and DMD.

Our main contributions are the development and evaluation of binary classifiers that differentiate between healthy volunteers and DMD or SMA patients as a proof-of-concept study as well as a three-class classifier (healthy control, DMD, SMA) for a differential diagnosis. As DMD and SMA are difficult to differentiate in early stages, an additional indicator could speed up the diagnostic process. We compare the use of US and PA images with different wavelengths and spectrally unmixed signals (SUS) as input to the network for classification. Furthermore, we analyse the performance of the classifiers across age groups/severity and identify next steps to provide clinicians with rapid diagnostic support for muscular diseases using PA imaging.

\section{Materials and methods}
In the following section, we will shortly describe the data sets, image preprocessing and the experimental setup for the two-class and three-class problem to differentiate between DMD, SMA and healthy volunteers.

\subsection{Datasets}
Image data from two studies considering both DMD and SMA were provided by the Department of Pediatrics and Adolescent Medicine at the University Hospital of Erlangen.
In these studies PA raw data were generated from 10 patients with DMD, 10 patients with SMA and 20 age-matched controls using a handheld PA system (MSOT Acuity Echo, iThera Medical GmbH, Munich, Germany). In total, the dataset consists of 10\,683 valid image frames per single-wavelength (WL) or per SUS. A WL or SUS is used for visualization of different tissue types. Each task was performed with four different input images, i.e., US images, two different WL images (800$\ts$nm, 920$\ts$nm) and a SUS image (Collagen). At this stage no raw data but only preprocessed data as provided by the vendor software were available, and combining different WL/SUS images did not provide a benefit within this setup. 
 
For DMD, which usually worsens with age, the data is divided into age groups which correlate with the progress of the disease. Three patients aged 5--6, five patients aged 7--8 and two patients aged 9--10 were examined. For SMA, the data is divided by the type of SMA. Patients with type SMA 1 exhibit more obvious symptoms than patients with SMA 2 or 3. For SMA, two patients with SMA 1, four patients with SMA 2 and four patients with SMA 3 were considered. For each class, the data is split patient-wise into training, validation and test set with ratio 0.5, 0.2, 0.3 to represent each severity stage of DMD and SMA in the test set. As the selection of the input image and network architecture was highly exploratory, we opted not to use cross-validation.

\subsection{Pre-processing}
To obtain the required PA images, the provided raw data was processed with the vendor specific-software to obtain four images containing one WL/SUS each. Furthermore, they were resized to 224$\times$224$\ts$px.

\subsection{Experimental setup}
A VGG16 network was trained for each classification task (control vs. DMD: 2-class DMD, control vs. SMA: 2-class SMA, control vs. DMD vs. SMA: 3-class) with each of the four different input image types separately. We then used the validation set to determine the best performing WL/SUS for each task for further investigation. For both the two-class and three-class classification tasks, a weighted cross entropy loss was used as loss function to counteract class imbalance. Stochastic gradient descent (SGD) with momentum was set as optimizer. VGG16 was pre-trained on the ImageNet dataset and all weights were frozen except for the last layer~\cite{ref-9}. Here, random augmentation techniques were applied to improve the training process (random brightness, noise, saturation, rotation and left-right flipping). Finally, the data was normalized to match ImageNet statistics. For the three-class problem, Gradient-weighted Class Activation Mapping (Grad-CAM) was used to highlight regions relevant for the network predictions from VGG16 and further investigate the network prediction. The resulting images were evaluated jointly with a pediatrician (F.K.) experienced in working with US and PA images of muscular tissue.

\section{Results}
As discussed in the previous section, a separate neural network was trained for each input image type and each classification task. During validation, US showed the best accuracy across all tasks with 0.96 (2-class DMD), 0.83 (2-class SMA) and 0.91 (3-class). Out of the PA images, 920$\ts$nm provides the best validation accuracy for 2-class DMD (0.95) and the 3-class case (0.83), whereas the 800$\ts$nm image showed the best accuracy for 2-class SMA.

For the final evaluation on the test set, we use the VGG16 networks trained with US and best performing PA image on the validation set, i.e., the 920$\ts$nm WL for the 2-class DMD and the 3-class case, and the 800$\ts$nm WL for the 2-class SMA. Table~\ref{table-test_us_measures} shows the final evaluation results on unseen images. All tasks show high accuracies ranging from 0.86 to 1.00 with better performance for the networks with US input, but also discriminative power from the networks with PA input.

\begin{table}[t]
	\centering
	\caption{Accuracy, recall, precision, AUC and F1 score for unseen test data for both 2-class and the 3-class tasks using US and the best PA WL/SUS as input and VGG16 as network architecture.}
	\begin{tabular*}{\textwidth}{l@{\extracolsep\fill}llllll}
		\hline
		&Input & Accuracy & Recall & Precision & AUC & F1 \\
		\hline
		\multirow[t]{2}{*}{2-class DMD}&US & 0.95 & 0.92 & 0.97 & 0.95 & 0.95 \\
		&920$\ts$nm & 0.90 & 0.82 & 0.96 & 0.90 & 0.90 \\
		\multirow[t]{2}{*}{2-class SMA}& US & 1.00 & 1.00 & 0.99 & 0.99 & 0.99 \\
		&800$\ts$nm & 0.86 & 0.97 & 0.69 & 0.89 & 0.86 \\
		\multirow[t]{2}{*}{3-class DMD}& US & 0.94 & 0.87 & 0.98 & 0.94& 0.92 \\
		&920$\ts$nm & 0.91 & 0.86 & 0.95 &0.91&  0.90 \\
		\multirow[t]{2}{*}{3-class SMA}& US &0.94 & 1.00 & 0.99 &0.94 & 0.99\\
		&920$\ts$nm & 0.91 & 0.59 & 0.93 & 0.91 &  0.72 \\
		\hline
	\end{tabular*}
	\label{table-test_us_measures}
\end{table}

\begin{figure}[b]
	\centering
	\setlength{\figwidth}{0.42\textwidth}
	\begin{subfigure}{\figwidth}
	    \includegraphics[width=\textwidth]{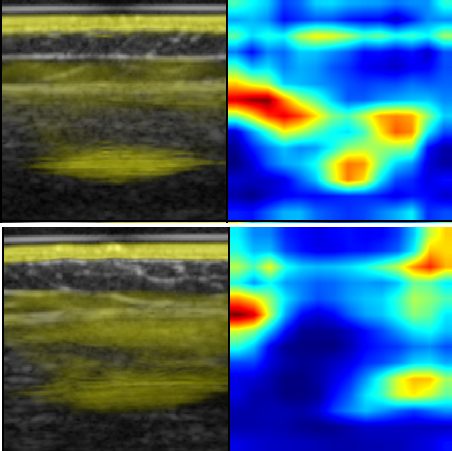}
	    \caption{SMA.}
	\end{subfigure}
	\hfill
	\begin{subfigure}{\figwidth}
	    \includegraphics[width=\textwidth]{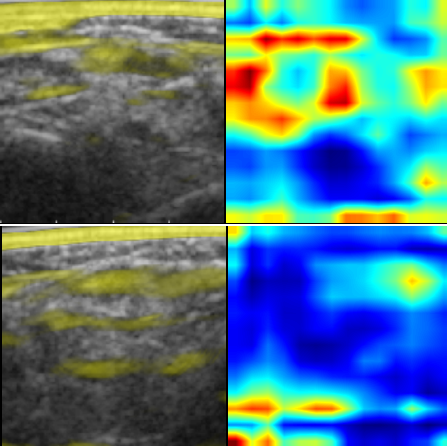}
	    \caption{DMD.}
	\end{subfigure}
	\hfill
	\begin{subfigure}{\figwidth}
	    \includegraphics[width=\textwidth]{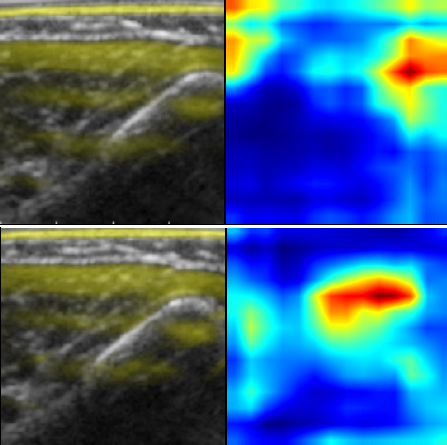}
	    \caption{Healthy.}
	\end{subfigure}
	\caption{Grad-CAM results for (a) SMA, (b) DMD and (c) healthy. Left: 920$\ts$nm image (yellow) overlayed on US image, right: heatmap of activated gradients. Highly activated regions are depicted in red and localize class discriminative features. Top row is correctly classified, bottom row is wrongly classified.}
	\label{figure-dmd_sma_US_grad}
\end{figure}

\begin{table}[t]
	\centering
	\caption{Percentage of frames classified correctly as DMD or SMA (true positive rate, 3-class case) for three DMD and three SMA patients broken down by different disease states using US and 920$\ts$nm images as input. Left: DMD, ages 5,7 and 10. Right: SMA type 1, 2 and 3.}
	\begin{tabular*}{\textwidth}{l@{\extracolsep\fill}llllll}
		\hline
		&  \multicolumn{3}{l}{DMD} & \multicolumn{3}{l}{SMA}\\
		\hline
		&  Age 10 & Age 7 & Age 5  & Type 1 & Type 2 & Type 3 \\
		US & 97\ts\% & 98\ts\% & 70\ts\%& 100\ts\% & 100\ts\% & 100\ts\%  \\
		920$\ts$nm & 90\ts\% & 99\ts\% & 69\ts\% & 100\ts\% & 49\ts\% & 31\ts\% \\
		\hline
	\end{tabular*}\hfill\quad
	\label{table-sma_dmd_results_tp_fn}
\end{table}
Table~\ref{table-sma_dmd_results_tp_fn} shows the true positive rate broken down by disease severity for all frames from the three DMD and SMA patients of the test set, respectively, using US and 920$\ts$nm input images. As expected, we mainly see errors in patients which are in an earlier stage of disease progression, i.e., DMD at age 5 and SMA type 3. While the error rates are similar for DMD across US and 920$\ts$nm, we see a considerably larger difference in errors for SMA between the input types. Table~\ref{table-sma_dmd_results_conf_matrix} shows the confusion matrix for the 3-class case using US and 920$\ts$nm images.
\begin{table}[t]
	\centering
	\caption{Relative confusion matrix for US and 920$\ts$nm of the 3-class case comparing network prediction and ground truth (GT) labels. Diagonal elements shows percentage of correctly classified frames.}
	\begin{tabular*}{\textwidth}{l@{\extracolsep\fill}lllllll}
		\hline
		&&  \multicolumn{6}{c}{Predicted} \\
		\hline
		& & \multicolumn{3}{l}{US} & \multicolumn{3}{l}{920$\ts$nm}\\
		& & Healthy & DMD & SMA  & Healthy & DMD & SMA \\
		\multirow[c]{3}{*}{GT}&Healthy & 98.6\ts\% & 1.3\ts\% & 0.1\ts\% & 97.3\ts\% & 2.3\ts\% & 0.4\ts\% \\
		&DMD& 12.5\ts\% & 87.5\ts\% &0\ts\%& 14\ts\% & 86\ts\% & 0\ts\%  \\
		&SMA& 0\ts\%& 0\ts\% & 100\ts\%& 33.6\ts\% & 7.1\ts\% & 59.3\ts\%  \\
		\hline
	\end{tabular*}\hfill\quad
	\label{table-sma_dmd_results_conf_matrix}
\end{table}
Figure~\ref{figure-dmd_sma_US_grad} shows the activated gradients for VGG16 in layer 29 of six different 920$\ts$nm WL images. One example each for SMA, DMD and healthy control which is correctly classified and one example each which is wrongly classified is exhibited. The visualization using Grad-CAM shows activated gradients for correctly classified SMA in the central muscle area. In the wrongly classified scan, no central activation is visible. For correctly classified DMD, strong activations are visible in more superficial muscular areas. For the correctly classified healthy scan, no central activations are present.

\section{Discussion}
In this paper, we show initial results for the classification of neuromuscular diseases based on US and PA images for pediatric patients suffering from DMD and SMA. These initial results on a small dataset are promising across the different settings and tasks. While US imaging provides slightly better quantitative classification results compared to PA input images in our study, clinical observations indicate the potential for an added value by PA images as it may offer finer gradation than US alone. Specifically, it is able to enrich the imaging data with functional information beyond the morphological structures visible in US. As we could not demonstrate a benefit in preliminary experiments yet, further research will have to investigate the potential of combining PA and US images and the limitations of this strategy under a more careful preprocessing scheme.

The dataset in the current study was acquired from a small study population with a comparatively simple network architecture and further evaluation has to be performed to validate our findings. In preliminary experiments, other architectures such as AlexNet, ResNet and EfficientNet-B0 were also considered and performed on par or slightly lower compared to VGG16. An evaluation of this performance difference will be interesting to discuss in future work. Due to the small dataset, we opted for a frame-wise evaluation of the classification. We expect that the classification performance can be increased further if multiple frames or scans of one patient are considered, e.g., via majority voting. On a larger dataset, we further see considerable potential for using multiple input images jointly for a classification to combine complementary information across US and PA, i.e., combining morphology and functional information.
This also translates to other applications such as inflammatory diseases where increased blood circulation in the injured region may be identified in the PA images whereas morphological changes can be identified in US images.

Our analysis shows a connection between the number of correctly classified scans and age for DMD, and correctly classified frames and subtype for SMA. This matches well with the fact that older patients in the case of DMD also exhibit more distinct clinical symptoms~\cite{ref-1} and that clinical symptoms vary in severeness in the different SMA subtypes~\cite{ref-10}. It can therefore be postulated that the disease progression also has clear impact both on the measured US and PA signal. US and PA could potentially be used to monitor treatment process and medication response to support personalized treatment planning and assessment if further research confirms predictability of severity for each disease. Despite the slightly lower performance of PA compared to US, PA's inherent properties will make it an interesting choice for future applications.

% \begin{disclaimer}
% 	Für Ausschlussklauseln nutzen Sie bitte die \texttt{disclaimer} Umgebung.
% \end{disclaimer}

% \begin{acknowledgement}
% 	Bitte schreiben Sie Ihre Danksagungen innerhalb der \texttt{acknowledgement} Umgebung. 
% \end{acknowledgement}

% Dieser Befehl erzeugt die Bibliographie mit Hilfe der Einträge der .bib-Datei. 
% Entfernen Sie diesen nur, wenn Sie keine Bibiographie benutzen. 
\printbibliography

\end{document}